# Experimental and Physics Prospects at ATLAS and CMS – 2011 and Beyond

Allan G. Clark
*DPNC, Université de Genève, 1211 Genève 4, Switzerland*

on behalf of the ATLAS and CMS Collaborations

The ATLAS and CMS experiments have collected data at the CERN Large Hadron Collider (LHC) since December 2009, and with a collision energy √s=7 TeV since March 2010. Both detectors work remarkably well at this early stage of operation, and several physics analyses have already been published. It is currently expected that an integrated luminosity of ~1 fb$^{-1}$ will be collected before a 15 month shutdown from early 2012, and that up to 10 (100) fb$^{-1}$ will be collected at a collision energy at or near √s=14 Tev by the end of 2013 (2016). Taking account of the outstanding Tevatron results obtained so far, a perspective is given of key physics measurements using data at the end of 2011 and beyond.

## 1. INTRODUCTION

The ATLAS [1] and CMS [2] experiments at the CERN Large Hadron Collider (LHC) [3] are the two largest and most polyvalent particle physics detectors ever built. Following the first proton-proton collisions at a collision energy of √s = 7 TeV on March 30, 2010, the detectors have collected data with high efficiency in a new energy domain. This talk discusses some key precision measurements and discovery searches that will be made by the experiments in 2011 and beyond, in the context of existing precision measurements from the Tevatron experiments.

A primary reason for the construction of the LHC and its experiments was (is) the discovery of the Higgs particle. Although at lower collision energy (√s = 1.96 TeV) the CDF and D0 experiments at the Tevatron Collider have made direct experimental inroads on this search with 95% confidence level exclusion of the Standard Model (SM) Higgs in the mass range 158 < $m_H$ < 175 GeV [4]. When combined with the LEP limit $m_H$ > 114.4 GeV [5] and the preferred value within the context of the SM of $m_H = 89^{+35}_{-26}$ GeV [6] resulting primarily from precision measurements of $m_{top}$ [7] and $m_W$ [8] at the Tevatron, it is useful to compare with the expected early data from ATLAS and CMS. Following a full understanding and optimization of the detector performance, the accuracy of important SM measurements such as $m_t$ and $m_W$ can potentially be improved.

The cross-sections for high $E_T$ processes and high mass particle production increase significantly when comparing LHC p-p collisions at √s = 7 TeV with Tevatron data: typically a factor > 20 for $t\bar{t}$ production, ~25 for SM Higgs production with $m_H$ =115 GeV, and up to 100 for hypothetical particle production at $m_X$ ~ 0.5 – 1 TeV. On the basis of the expected LHC performance and assuming optimal detector performance, the ATLAS and CMS experiments should rapidly improve on existing searches for the Higgs particle and for possible experimental signatures "Beyond the Standard Model (BSM)".

In this talk, we discuss a small number of precision measurements and search channels, assuming a data sample of integrated luminosity $L_{int}$ ≤ 1 fb$^{-1}$ at √s = 7 TeV and $L_{int}$ = 10 fb$^{-1}$ at √s = 14 TeV. In Section 2, the expected LHC performance is briefly noted, and both the existing and optimal measurement precisions of the different ATLAS and CMS sub-detectors are summarized. The physics potential is then explored for precision SM measurements (Section 3). The short and medium term prospects for SM Higgs identification and measurement are summarized in Section 4, and the search for signatures characteristic of new BSM physics using as examples deviations of the di-jet mass spectrum from QCD expectations, the search for heavy gauge bosons (W' or Z') and inclusive super-symmetric (SUSY) particle production (Section 5). Short conclusions are made in Section 6. The slides as presented give more detail on a number of other BSM processes.

## 2. PERFORMANCE OF THE LHC MACHINE AND EXPERIMENTS

### 2.1. The LHC Machine – Expected Evolution of Luminosity

As described in this conference [9], the LHC machine luminosity has rapidly increased, with a peak luminosity of $L_{peak} \sim 1.5 \times 10^{31}$ cm$^{-2}$s$^{-1}$, and bunch trains of up to 50 bunches. An integrated luminosity $L_{int} = 2.8$pb$^{-1}$ has been accumulated. It is expected that the peak luminosity will be increased to beyond $2 \times 10^{32}$ cm$^{-2}$s$^{-1}$ in 2010 and that an integrated luminosity $L_{int} \sim 1$fb$^{-1}$ will be accumulated before a shutdown in early 2012. There will also be 1-month Heavy Ion (HI) runs at the end of 2010 and 2011. There are 3 major shutdowns currently foreseen: a 15-month shutdown in 2012 to insert helium relief valves and to complete the magnet interconnection repairs enabling collision energies at or near 14 TeV; a 12-month shutdown in 2016 to integrate the Linac4 in the injector chain and upgrade the CERN PS booster to allow operation at luminosities $L_{peak} \sim 2 \times 10^{34}$ cm$^{-2}$s$^{-1}$; and an 18-month shutdown in 2020 for major machine and detector modifications needed for high luminosity running ($L_{peak} \geq 5 \times 10^{34}$ cm$^{-2}$s$^{-1}$). A realistic estimate of the integrated luminosity expected until 2020 is shown in Table 1. Most comments in this talk will consider the period until the end of 2013: $L_{int} \sim 1$ fb$^{-1}$ at $\sqrt{s} \sim 7 - 8$ TeV, and $L_{int} \sim 10$ fb$^{-1}$ at $\sqrt{s} \sim 14$ TeV.

Table I: Expected data samples at the CERN LHC as of August 2010

| End date | Collision Energy (TeV) | $L_{peak}$ (cm$^{-2}$s$^{-1}$) | $L_{int}$(fb$^{-1}$) |
|---|---|---|---|
| 2010 | 7.0 | $10^{32}$ | 0.07 |
| 2011 | ~8 (?) | $1.9 \times 10^{32}$ | 1.04 |
| 2013 | 13 – 14 | $2.6 \times 10^{33}$ | 9.2 |
| 2014 | 14 | $5 \times 10^{33}$ | 30.0 |
| 2015 | 14 | $10^{34}$ | 66.3 |
| 2017 | 14 | $10^{34}$ | 118.1 |
| 2018 | 14 | $1.7 \times 10^{33}$ | 216.9 |
| 2019 | 14 | $2.2 \times 10^{34}$ | 335.8 |

## 2.2. Expected Performance Evolution of the ATLAS and CMS Detectors

The data from ATLAS [10] and CMS [11] are already of a quality beyond what might have been expected at this early stage. However, the timing, calibration and alignment optimization of the detectors remains to be completed, and detailed instrumental comparisons of the data with design expectations are ongoing. In most cases, only a few pb$^{-1}$ of data are required to optimize the calibration and alignment constants, but initially these must be repeated frequently to understand the performance stability. An exception is the electron scale linearity and uncertainty at high $p_T$, where large samples of $Z \to e^+e^-$, $Y \to e^+e^-$ and $J/\Psi \to e^+e^-$ decays are desirable.

a) Issues for the Inner Tracker (ID) include an optimization of the track reconstruction efficiency, and an optimization of the pixel or silicon strip alignment to achieve in a stable way the design primary and secondary vertex resolution and the track momentum resolution ($\sigma/p_T$ (GeV)). Also important is an accurate solenoidal magnetic field ($B_{sol}$) calibration. The r-φ (bending plane) alignment tolerances are already at the < 15-20 μm level, and the effort is towards the understanding of systematic effects such as module distortions, and minor mode redundancies. The required r-φ alignment tolerances are < 7 (10) μm for the pixels (silicon strips). This results in a design resolution in the central region of $3.8 \times 10^{-4}$ $p_T \oplus 0.015$ for ATLAS and $1.5 \times 10^{-4}$ $p_T \oplus 0.015$ for CMS, the difference being mostly due to the 4 Tesla CMS field. The tracker material of both experiments must be understood to better than ±1% (the current knowledge is at the 5-10% level, with the detailed understanding of services being a key issue).

b) The electromagnetic calorimeter issues include the resolution, and linearity, as well as the overall energy scale, the channel-to-channel variation, and the calibration stability. The design performance is substantially improved with respect to previous detectors. The expected uncertainty on the scale is ±0.1% for each experiment, with ~ ±0.5% channel-to-channel uncertainty. The currently quoted combined values are already ±3% for ATLAS and ±1.5% for CMS. When large $Z \to e^+e^-$ samples are available, the energy scale uncertainty in the vicinity of Z will be significantly less than quoted above. A combined uncertainty of ±7% is currently quoted for each experiment for the hadron calorimeters, while the final uncertainty is expected to be < ±2%.

c) For the muon spectrometers, the required bending plane alignment is < 20 – 30 μm (ATLAS) and < 100 – 200 μm (CMS) because it is not used in stand-alone mode. The reconstructed $Z \to \mu^+\mu^-$ peaks of each experiment already demonstrate the excellent progress in understanding the detector.

The measurement precision of reconstructed physics objects is what will ultimately determine the ability of ATLAS and CMS to improve on existing published data for $m_{top}$ and $m_W$. For cross-section measurements, a luminosity uncertainty of $\delta L_{int} < \pm 5\%$ is expected in the near future: the current value of ±11% is dominated by the beam current measurement uncertainty.

## 3. STANDARD MODEL MEASUREMENTS

The importance of measuring the production and properties of major known SM processes cannot be underestimated, firstly because of their intrinsic physics interest in confirmation or otherwise of the SM, and secondly because they provide major background contributions for BSM processes. Until such time as the Higgs particle is directly observed, arguably the most important precision mass measurements are $m_W$ and $m_{top}$ since they provide an indirect constraint on $m_H$ within the framework of the SM.

## 3.1. Measurement of $\sigma_W$ and $m_W$ at ATLAS and CMS

Initial measurements of $\sigma_W$ and $\sigma_Z$ in their leptonic decay modes have been shown at this conference [12,13]. With a data sample $L_{int}$ = 1 fb$^{-1}$ (> 250K $Z \to l^+l^-$ events and > 2.5M $W^\pm \to l^\pm \nu_l$ events), systematic uncertainties will dominate the accuracy of these important QCD measurements, in particular the luminosity uncertainty ($\delta L_{int} < \pm 5\%$), the uncertainty of background subtraction and the effect of different Monte Carlo generators. The Tevatron experiments report uncertainties of ~ ±2% for the background and modeling uncertainties, and that uncertainty should be achievable at ATLAS and CMS.

The measurement of $m_W$ is again systematics dominated. Issues include: the compatibility of the $e^\pm$ and $\mu^\pm$ decay channels; the compatibility of the Monte Carlo modeling of the $m_T^{l\nu}$, $E_T^l$ and missing $E_T$ ($E_T^{miss}$) observables; a full understanding in $\eta-\phi$ of the tracker material (to be known to $< \pm 1\%$); and the W-mass energy calibration (as noted above, transferred from $m_Z$ using a measurement of $m_W/m_Z$ and therefore requiring a high Z statistics). The current world average $m_W = 80.399 \pm 0.023$ GeV [8] includes a best individual measurement by the D0 experiment (~500K events, L ~1 fb$^{-1}$) [14] of $m_W = 80.401 \pm 0.021(stat) \pm 0.038(sys)$ GeV. In this case, the systematics is dominated by the energy scale uncertainty (±35 MeV). The 2011 data samples from CMS and ATLAS should eventually allow a $m_W$ measurement of equivalent precision. With the 2013 data sample (L ~10 fb$^{-1}$, √s=14 TeV) the expected ATLAS and CMS precision is $\delta m_W(stat)$ ~ ±2 MeV and $\delta m_W^{\mu,e}(sys)$ ~ ±6(7) MeV [1,2].

## 3.2. Measurement of $\sigma_{t\bar{t}}$ and $m_t$ at ATLAS and CMS

As shown at this conference [15,16], $t\bar{t}$ event samples have been identified at the LHC, with L ~ 1pb$^{-1}$. With $\sigma_{t\bar{t}}^{7\,TeV}$ ~ 20–25 x $\sigma_{t\bar{t}}^{1.96\,TeV}$, $t\bar{t}$ data samples of ~25 K single-lepton and ~3 K di-lepton events are expected for each experiment using the 2011 data set. A measurement of $\sigma_{t\bar{t}}$ (and $\sigma_{W+n(jet)}$) in different decay channels is an essential demonstration of the detector performance and will allow an accurate modeling of the background for many BSM signatures. The current CDF measurement accuracy (quoted with respect to Z,γ production to avoid the luminosity uncertainty) of ±4.1%(stat), ±4.5%(sys) and ±2%(theory) [17] should be significantly improved even with the 2011 data sample.

A measurement of $m_t$ is the second key ingredient in providing indirect limits on $m_H$. The combined Tevatron measurement of $m_t = 173.3 \pm 0.6(stat) \pm 0.9(sys)$ [18] is an outstanding achievement. In the case of CDF, the major systematic uncertainties include the statistical component of the jet energy scale (±0.46 GeV), the effect of background subtraction (±0.23 GeV), and an understanding of the shower model and color recombination (±0.56 GeV). A competitive LHC measurement is statistically feasible using the 2011 data sample, but a detailed understanding of the detector, in particular the jet energy scale uncertainty, will take time and focus. For $L_{int}$ ~ 10fb$^{-1}$ at √s=14 TeV (the 2013 data sample), the available statistics will enable tight kinematical selections that reduce some systematic uncertainties. Also, the available statistics will allow a competitive measurement using the di-lepton signature.

## 4. STANDARD MODEL HIGGS PRODUCTION AT THE LHC

As already noted, the precision mass measurements of $m_W$ and $m_t$ at the Tevatron leading to a best-fit SM value $m_H = 89^{+35}_{-26}$ GeV [6], as well as direct experimental searches [4,5] that provide a 95% exclusion range for the SM Higgs of $m_H < 114.4\ GeV$ and $158 < m_H < 175\ GeV$, restrict the discovery window (barring surprises). However, with $\sigma_H^{7\ TeV} \sim 25\ \mathrm{x}\ \sigma_H^{1.96\ TeV}$ even in the low mass range ($m_H \sim 115 - 120$ GeV), the expected 2011 data sample will exceed that available at the Tevatron. The dominant (80%) LHC production process is gluon fusion. However, particularly in the low-mass region, the QCD background contributions are higher at the LHC. Figure 1 shows branching ratio for different SM Higgs decays as a function of $m_H$.

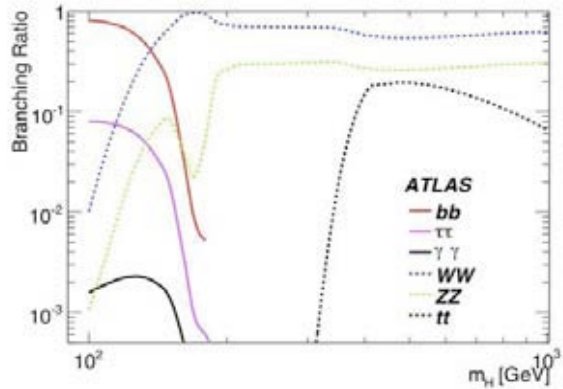

Figure 1. The SM Higgs branching ratio as a function of the Higgs mass [1], see also A. Djouadi [19]. Note that these branching ratios may be substantially modified in BSM scenarios.

a) In the range $115 < m_H < 140\ GeV$, the $H \to b\bar{b}$ decay mode dominates. The decay $H \to \tau^+\tau^-$ is also non-negligible. The ability to identify these decays from the QCD background, possibly using forward quark tags as in the $t\bar{t}$ and vector boson fusion processes, must still be proven (these decay modes are more favorable at the Tevatron and have been included in recent CDF and D0 analyses). The most promising decay signature (see below) is the reconstruction of low rate $H \to \gamma\gamma$ events.

b) For $m_H > 130\ GeV$, detection via the decays $H \to WW^{(*)}$ and $H \to ZZ^{(*)}$ becomes feasible. With increasing mass, the $H \to WW^{(*)}$ event rate decreases and the challenge (as with the $H \to ZZ^{(*)}$ leptonic signature) becomes statistical; at lower mass, important SM background signatures from $t\bar{t}$, single t-quark production, WW production and W+$n_{jet}$ production must be understood. The analyses have so far been largely cut-based because of the early stage of the experiment [20,21]. However, as with the Tevatron searches, sophisticated multi-dimensional analysis techniques can be expected rapidly. Selecting 2 isolated high-$p_T$ leptons with associated $E_T^{miss}$, and making use of the lepton angular correlations expected with $W^\pm$ decay, the ATLAS experiment (Figure 2) expects to be able to exclude (95%) the range $140 < m_H < 185\ GeV$ using the expected 2011 data set. A similar analysis power is expected from CMS; each result improves on the current Tevatron limit, more powerful analyses are being developed, and the results can be combined.

c)  The $H \to \gamma\gamma$ signature is very actively studied by ATLAS and CMS for lower $m_H$ (and $\gamma\gamma$ final states are also relevant to many possible BSM signatures, for example Kaluza Klein graviton searches). It is measured as a narrow mass peak superimposed on a dominant $\gamma\gamma$ background of irreducible QCD processes, and a reducible background dominantly from (γ+jet), Drell-Yan and QCD multi-jet processes. Because of the small signal, attempts are made to recover converted photons and a key issue in both experiments is the understanding of the Inner Detector material distribution, actually known at the ± 5-10% level, but which must ultimately be known to < ±1%. The ATLAS and CMS detectors have complementary strengths and weaknesses [1,2]: while CMS has a better calorimetric energy resolution (subject to calibration), ATLAS has a superior directional photon measurement, allowing the reduction of background from multiple vertices etc. As noted at this conference, ATLAS and CMS have already isolated clean single photon signals [22,23]. Using a preliminary ATLAS analysis [24] at $\sqrt{s}$ = 7 TeV with $L_{int}$ = 1 fb$^{-1}$, Table 2 shows the expected signal and background statistics.

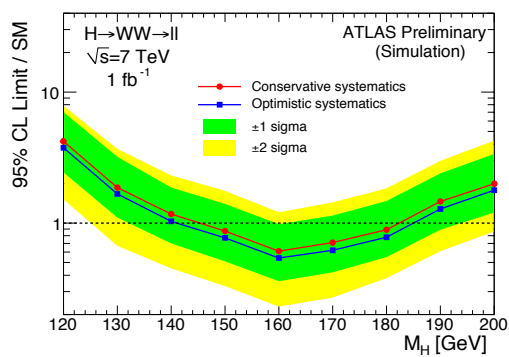

Figure 2: Expected 95% CL upper limits of the Higgs boson production cross-section normalized to the SM cross-section. The green and yellow bands represent the range in which the limit is expected to lie, depending upon the data. The effect of conservative and optimistic systematic uncertainties is indicated.

Table 2: Expected data samples at the CERN LHC as of August 2010

| ATLAS: Expected signal and background ($\sqrt{s}$ = 7 TeV, $L_{int}$ = 1 fb$^{-1}$) | | | |
|---|---|---|---|
| $m_H$ (GeV) | 110 | 120 | 130 |
| Background | 5540 (irreducible) 2950 (reducible) range 100 < $m_H$ < 150 GeV | | |
| Signal | 12.6 | 12.8 | 13.0 |
| $L_{int}$ (95% excl.) | 5.8 fb$^{-1}$ | 4.6 fb$^{-1}$ | 5.2 fb$^{-1}$ |

Combining the $WW^{(*)}$, $ZZ^{(*)}$ and $\gamma\gamma$ channels using the 2011 data set, ATLAS and CMS quote the following expected 95% exclusion limits: 135 < $m_H$ < 188 GeV (ATLAS) and 145 < $m_H$ < 190 GeV (CMS). These limits are expected to further improve[1], but are already individually competitive with Tevatron data. In the longer term, Figure 4 from the CMS experiment shows that with $L_{int}$ = 10 fb$^{-1}$ at $\sqrt{s}$=14 TeV (2013), a 5σ discovery can be envisaged over an extended mass range.

---

[1] Subsequent to this conference, ATLAS has presented updated expectations indicating an improved Higgs sensitivity [25].

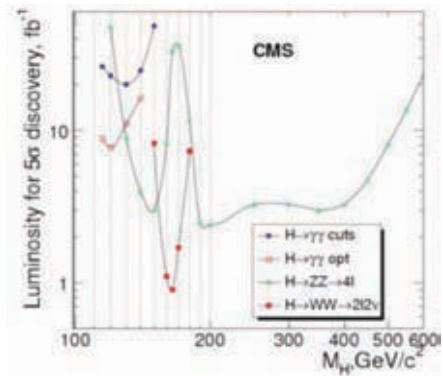

Figure 3: The integrated luminosity needed for the 5σ discovery of the inclusive Higgs boson production $pp \rightarrow H + X$ with the Higgs boson decay modes $H \rightarrow \gamma\gamma$, $H \rightarrow ZZ^{(*)} \rightarrow 4l$ and $H \rightarrow WW^{(*)} \rightarrow 2l2\nu$ [2].

## 5. SEARCHING BEYOND THE STANDARD MODEL (BSM)

A wide range of BSM searches are already advanced in both the ATLAS and CMS experiments and several are presented at this conference [26,27]. With $L_{int} < 1$ pb$^{-1}$, some limits for high mass processes are already competitive with Tevatron data. The 2011 data set will supersede most existing BSM searches because of the new collision energy scale. The breadth of these searches is beyond the scope of this paper, and we restrict our comments to a number of "benchmark" BSM signatures. BSM searches for resonant di-photon structures have been alluded to above.

### 5.1. Deviation from single- and di-jet QCD distributions

The ATLAS experiment has used a data sample $L_{int} = 315$ nb$^{-1}$ to search for resonant deviations from the di-jet mass spectrum, $d\sigma/dm_{JJ}$. Assuming the MRST2007 modified leading-order parton distributions, and an excited quark model, they report a 95% exclusion limit $400 < m_{q^*} < 1265$ GeV, to be compared with the existing limit of 870 GeV [28,29]. Figure 4 shows the expected CMS 5σ discovery potential at the end of the √s=7 TeV 2011 run, for several theoretical models including excited quarks [30]. Using a data sample of $L_{int}=10$ fb$^{-1}$ at √s=14 TeV (the expected 2013 data set), the 95% confidence $m_{q^*}$ exclusion limit is expected to reach $m_{q^*}$ ~ 4.4 TeV in each experiment if no discovery is made.

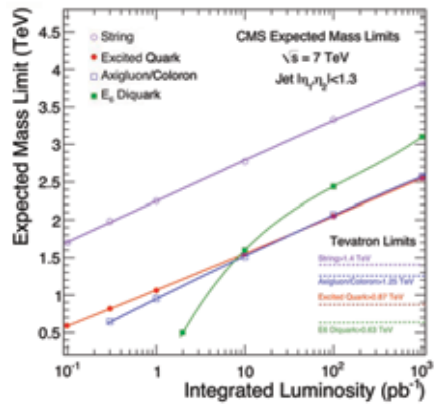

Figure 4: The CMS discovery potential for dijet resonances at 7 TeV as a function of $L_{int}$ for several popular models.

The CMS experiment has used the centrality ratio $R = N(|\eta| < 0.7)/N(0.7 < |\eta| < 1.3)$ to search for non-resonant di-jet deviations from QCD [31]. With a data sample $L_{int}$ = 120 nb$^{-1}$, and defining the deviations in terms of a contact interaction scale Λ, they obtain good agreement with QCD expectations and require with 95% confidence Λ > 1900 GeV. With $L_{int}$ ~ 4 pb$^{-1}$, this limit should increase to Λ > 3 TeV, improving on the existing Tevatron limit Λ > 2.8 TeV [32]. For $L_{int}$ = 10 fb$^{-1}$ at √s = 14 TeV, the 95% confidence exclusion limit on Λ is expected to be Λ ~ 15 TeV in each experiment. A 5σ observation of a deviation from QCD should be possible for Λ < 12 TeV.

### 5.2. Heavy Gauge Boson Production

High mass di-lepton states are expected in many models: gauge bosons in Grand Unified Theories (GUT), Technicolor, little Higgs models, Extra Dimension scenarios such as the Randall-Sundrum graviton, etc. Most searches assume a narrow resonance (typically 1% of the mass), but there is some dependence on the different models. Assuming a Z' coupling, Figure 5a) shows the expected exclusion limit in ATLAS for $Z' \to l^+l^-$ final states as a function of $L_{int}$ at √s = 7 TeV assuming SM couplings. Ultimately, exclusion (or discovery) of $m_{Z'}$ < 3 – 4 TeV can be envisaged. Using high-$p_T$ single lepton samples with associated $E_T^{miss}$ Figure 5b) shows the expected exclusion limit for W' couplings in 7 GeV data [30,33].

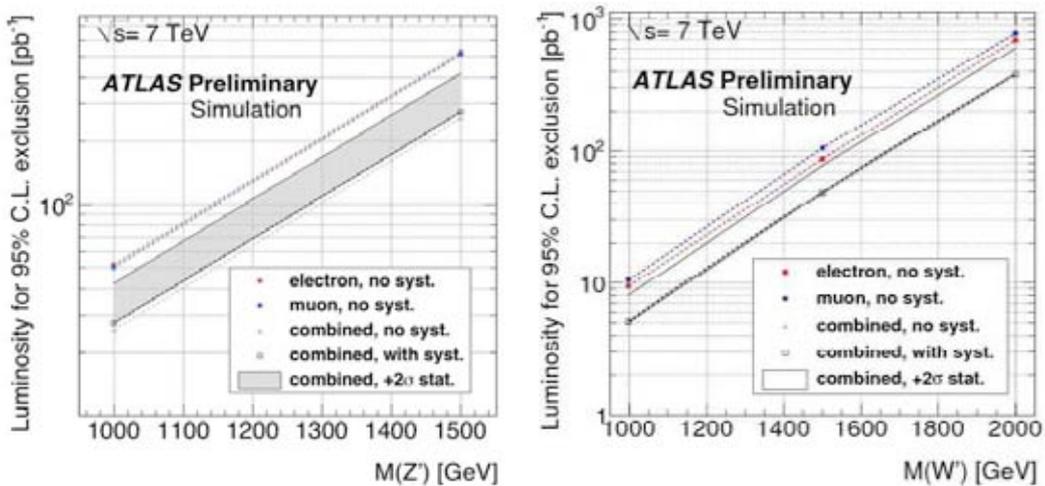

Figure 5: (Left) Integrated luminosity expected to allow a 95% CL exclusion of the SSM Z' model, as a function of M(Z'). (Right) 95% C.L. exclusion limits on the W' production cross section assuming the same branching ratio for all leptons.

### 5.3. The Search for Supersymmetric particles

A major incentive for the LHC construction has been to understand BSM processes that might enable extensions of the SM to be stable mathematically at high collision energies (the hierarchy and fine tuning problem). As discussed at this conference [34], a favored BSM strategy is Supersymmetry (SUSY), where every fermionic quark has a bosonic squark of spin 0 and the gauge bosons have a fermionic counterpart (chargino, gluino, etc). There is to date no direct experimental evidence for SUSY, but it has the advantage (in the case of symmetry breaking at energies on the TeV scale with conservation of a SUSY quantum number called R-parity) that the lightest neutral SUSY particle might be a dark matter candidate, as required by astronomical observations. The parameter space for Supersymmetry (SUSY) is vast, and in practice simplified production and decay schema are developed to allow a largely model-independent search strategy. The event signatures are normally characterized by large $E_T^{miss}$, and possibly by a high $p_T$ single lepton or di-lepton signature (for example the mSUGRA simplification) with associated jets. Variations of this simplistic model also allow di-photon production etc (for example the GMSB simplification).

One search strategy at ATLAS and CMS, as well as the Tevatron experiments, is to remain as model-independent as possible [26,27]. A detailed understanding of the SM contribution to the $E_T^{miss}$ distribution in events with associated jet production (and possibly as noted above an additional signature, for example a b-tag) is therefore important. Figure 6 shows early existing data from ATLAS for the $E_T^{miss}$, and a comparison from Monte Carlo events. There is no evidence so far for SUSY but deviations from the SM at large $E_T^{miss}$ would be a hint. Because of the increased collision energy, even the LHC data set from 2010 should provide limits on SUSY production beyond existing Tevatron limits.

Within the mSUGRA framework, the best existing limits are from D0 ($m_{gluino}$ > 308 Gev, and $m_{squark}$ > 379 GeV with 95% confidence level) [35]. Because of the increased collision energy, existing Tevatron limits should be attained with an integrated luminosity $L_{int}$ ~50 pb$^{-1}$. Figure 7 shows the expected exclusion limits for CMS and ATLAS with the 2011 data sample of $L_{int}$ ~1 fb$^{-1}$ [29,36]. For $\sqrt{s}$ = 14 TeV, the sensitivity is increased by a further factor of 20. A broad range of SUSY searches are in progress.

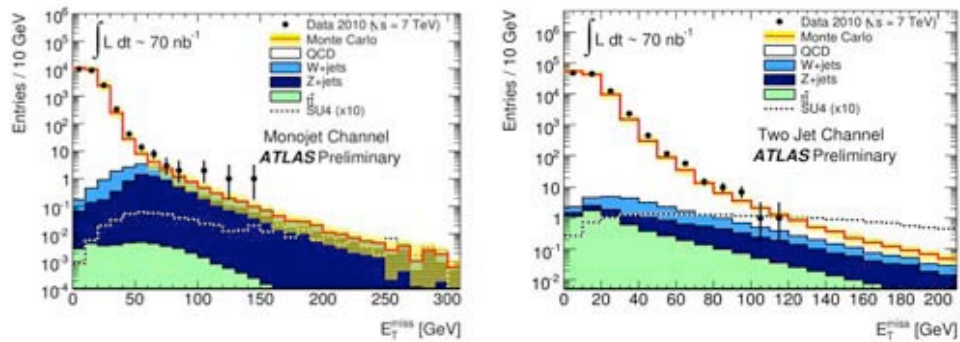

Figure 6: Distribution of $E_T^{miss}$ for events in the monojet (a) and di-jet (b) channels using an initial data sample $L_{int}$=70 nb$^{-1}$. Only the jet selection cuts have been applied.

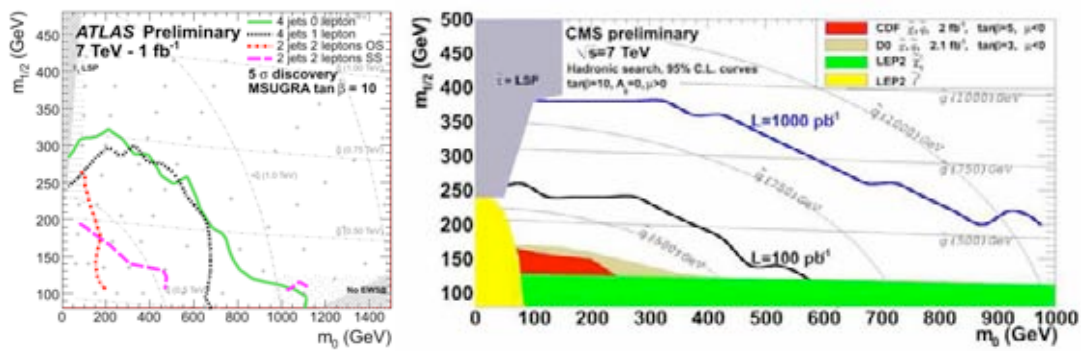

Figure 7: (Left) 5σ discovery reach in ATLAS as a function of $m_0$ and $m_{1/2}$ masses for tanβ = 10 mSUGRA scan for channels with 0, 1 and 2 leptons. (Right) 95% C.L. exclusion limits for the all-hadronic SUSY search in CMS, expressed in mSUGRA parameter space.

If a SUSY-like signal is established, the next questions will be: is it SUSY or some other BSM signal; if so what kind of SUSY; and what theoretical characteristics? It is fortunate (or perhaps naïve) that most BSM experimental signatures have similar final states. If a signal is established, the follow-up will be long, painstaking and tedious, but very exciting.

## 6. CONCLUSIONS

The ATLAS and CMS experiments at the CERN Large Hadron Collider have collected first data at $\sqrt{s}$ = 7 TeV. They are operating close to design specifications, and at the same time the LHC is making outstanding progress in its commissioning phase.

In this presentation we indicate some of the early results from ATLAS and CMS, and the expected physics potential with respect to existing experiments at the Fermilab Tevatron, after key data collection goals are reached: the collection of $L_{int}$ ~ 1 fb$^{-1}$ at a collision energy of $\sqrt{s}$ = 7 TeV (foreseen by the end of 2011), and the collection of $L_{int}$ ~10 fb$^{-1}$ at a collision energy of 14 TeV (foreseen by the end of 2013).

As the machine commissioning develops, the details of the experimental scenario will certainly change. However, the future physics potential at the LHC remains very encouraging, provided a detailed understanding of the ATLAS and CMS performance, as evidenced by a few key precision measurements such as $m_W$ and $m_t$, is developed over a relatively short period. Even on the very short time scale, data from the LHC will be competitive or superior to data from the Fermilab Tevatron, given the increased collision energy.

## Acknowledgements

The author wishes to thank the organizers of this outstanding HCP2010 Symposium, at a historic period of first LHC data taking. The author would also like to thank his colleagues on the ATLAS and CMS experiments for the opportunity to speak on their behalf and for their help in preparing this presentation.## References

[1]   ATLAS Collaboration, G. Aad et al., *The ATLAS Experiment at the CERN Large Hadron Collider,* JINST 3 (2008) S08063.
      ATLAS Collaboration, G. Aad et al., *Expected Performance of the ATLAS Experiment: Detector, Trigger and Physics,* arXiv:0901.0512 (2008), CERN-OPEN-2008-020

[2]   CMS Collaboration, R. Adolphi et al., *The CMS Experiment at the CERN LHC*, JINST 3 (2008) S08004

      CMS Collaboration, G. l. Bayatian et al., *Technical Design Report, Volme II: Physics Performance,* CERN LHCC 2006-081

[3]   *The CERN Large Hadron Collider: Accelerator and Experiments*, CERN, 2009 (eds. A. Breskin and R. Voss)

[4]   CDF and D0 Collaborations, Tevatron New Physics, Higgs Working Group, *Combined CDF and DØ Upper Limits on Standard-Model Higgs-Boson Production with up to 6.7 fb$^{-1}$ of Data*, arxiv:1007.4587 [hep-ex]

[5]   LEP Working Group for Higgs Boson Searches, *Search for the Standard Model Higgs boson at LEP*, Phys. Lett. B565, 61-75 (2003)

[6]   See http://lepewwg.web.cern.ch/LEPEWWG/

[7] CDF and D0 Collaborations, Tevatron Electroweak Working Group, *Combination of CDF and D0 Results on the Mass of the Top Quark*, arXiv:1007.3178v1 [hep-ex]

[8] CDF and D0 Collaborations, Tevatron Electroweak Working Group, *Updated Combination of CDF and D0 Results for the Mass of the W Boson*, $m_W$ at Tevatron, arXiv:0908.1374v1 [hep-ex]

[9] S. Myers, *Report on the LHC*, 35$^{th}$ International Conference on High Energy Physics, Paris July 2010, see http://www.ichep2010.fr/
R. Bailey, *Status of the LHC Machine* and *LHC Upgrades/Plans*, this conference.

[10] S. Zimmermann, for the ATLAS Collaboration, *Status of ATLAS*, this conference

[11] G. Tonelli, for the CMS Collaboration, *Status of CMS*, this conference

[12] C. Mills, for the ATLAS Collaboration, *W and Z Physics at ATLAS*, this conference

[13] K. Hahn, for the CMS Collaboration, *W/Z Physics at CMS*, this conference

[14] D0 Collaboration, V.M. Abazov et al., *Measurement of the W Boson Mass*, Phys. Rev. Lett. 103, 141801 (2009)

[15] G. Cortiana, for the ATLAS Collaboration, *First Results on Top Quark from ATLAS*, this conference

[16] F.-P. Schilling, for the CMS Collaboration, *Top Quark Studies with CMS*, this conference

[17] CDF Collaboration, T. Aaltonen et al., *Measurement of the Ratio $\sigma_{t\bar{t}}/\sigma_{Z/\gamma \to ll}$ and Precise Extraction of the $t\bar{t}$ Cross Section*, Phys. Rev. Lett. 105, 012001 (2010)

[18] T. Schwartz, for the CDF and D0 Collaborations, *Measuring Top Quark Properties at the Tevatron*, this conference

[19] A. Djouadi, *The Anatomy of Electroweak Symmetry Breaking. I: The Higgs Boson in the Standard Model*, arXiv:hep-ph/0503172v2 (2005)

[20] M. Schram, for the ATLAS Collaboration, *ATLAS Higgs Sensitivity for 1 fb$^{-1}$ at the LHC at 7 TeV*, this conference

[21] M. Sani, for the CMS Collaboration, *Prospects for Higgs Boson Searches with CMS*, this conference

[22] ATLAS Collaboration, *Evidence for Prompt Photon Production in pp Collisions at $\sqrt{s} = 7$ TeV*, ATLAS note ATLAS-CONF-2010-077

[23] CMS Collaboration, *Photon Reconstruction and Identification at $\sqrt{s} = 7$ TeV*, CMS note CMS PAS EGM-10-005

[24] ATLAS Collaboration, *ATLAS Sensitivity Prospects for Higgs Boson Production at the LHC Running at 7 TeV*, ATL-PHYS-PUB-2010-009

[25] ATLAS Collaboration, *ATLAS Sensitivity Prospects for 1 Higgs Boson Production at the LHC Running at 7, 8 or 9 TeV*, ATL-PHYS-PUB-2010-015

[26] P. Savard, for the ATLAS Collaboration, *Searches for New Physics with ATLAS*, this conference

[27] Sung-Won Lee, for the CMS Collaboration, *Searches for Physics Beyond the Standard Model at CMS*, this conference

[28] ATLAS Collaboration, G. Aad et al., *Search for New Particles in Two-Jet Final States in 7 TeV Proton-Proton Collisions with the ATLAS Detector at the LHC*, Phys. Rev. Lett. 105, 161801 (2010)

[29] CMS Collaboration, *Search for Dijet Resonances in the Dijet Mass Distribution in 7 TeV pp Collisions at CMS*, CMS Notes CMS PAS EXO-10-001 and EXO-10-010

[30] CMS Collaboration, *CMS physics reach for searches at 7 TeV*, CMS note 2010-008.

[31] CMS Collaboration, *Search for New Physics with the Dijet Centrality Ratio*, CMS Note CMS PAS EXO-10-002

[32] D0 Collaboration, V. Abazov et al., *Measurement of Dijet Angular Distributions at Searches for Quark Compositeness and Extra Spatial Dimensions,* Phys. Rev. Lett. 103 (2009) 191803
CDF Collaboration, T. Aaltonen et al., *Search for new particles decaying into dijets in proton- antiproton collisions at √s = 1.96 TeV*, Phys. Rev. D 79 (2009) 112002

[33] ATLAS Collaboration, *ATLAS sensitivity prospects to W' and Z' at 7 TeV*, ATL-PHYS-PUB-2010-077

[34] K. Matchev, *BSM Theory*, this conference

[35] D0 Collaboration, V. Abazov et al., Search for Squarks and Gluinos in Events with Jets and Missing Transverse Energy using 2.1 fb$^{-1}$ of $p\bar{p}$ Collision Data at √s = 1.96 TeV, Phys. Lett. B660, 449 (2008)

[36] ATLAS Collaboration, *Prospects for Supersymmetry discovery based on inclusive searches at 7 TeV centre-of-mass energy with the ATLAS detector*, ATL-PHYS-PUB-2010-010